\documentclass[11pt]{article}
\usepackage{graphicx,floatflt,amssymb,epsfig,epsf} 
\def\be {\begin{equation}}
\def\ee {\end{equation}}
\def\bea {\begin{eqnarray}}
\def\eea {\end{eqnarray}}

\def\g {\gamma}

\def\ztwo {{\cal Z}_2}

\def\stwo {\sqrt{2}}

\def\opcit(#1){ {\em op. cit.}, #1}
\def\etal {\em et al.}

\def\issue(#1,#2,#3){{\bf #1}, #2 (#3)} 

\def\APP(#1,#2,#3){Acta Phys.\ Polon.\ \issue(#1,#2,#3)}
\def\ARNPS(#1,#2,#3){Ann.\ Rev.\ Nucl.\ Part.\ Sci.\ \issue(#1,#2,#3)}
\def\CPC(#1,#2,#3){Comp.\ Phys.\ Comm.\ \issue(#1,#2,#3)}
\def\CIP(#1,#2,#3){Comput.\ Phys.\ \issue(#1,#2,#3)}
\def\EPJC(#1,#2,#3){Eur.\ Phys.\ J.\ C\ \issue(#1,#2,#3)}
\def\EPJD(#1,#2,#3){Eur.\ Phys.\ J. Direct\ C\ \issue(#1,#2,#3)}
\def\IEEETNS(#1,#2,#3){IEEE Trans.\ Nucl.\ Sci.\ \issue(#1,#2,#3)}
\def\IJMP(#1,#2,#3){Int.\ J.\ Mod.\ Phys. \issue(#1,#2,#3)}
\def\JHEP(#1,#2,#3){J.\ High Energy Physics \issue(#1,#2,#3)}
\def\JPG(#1,#2,#3){J.\ Phys.\ G \issue(#1,#2,#3)}
\def\MPL(#1,#2,#3){Mod.\ Phys.\ Lett.\ \issue(#1,#2,#3)}
\def\NP(#1,#2,#3){Nucl.\ Phys.\ \issue(#1,#2,#3)}
\def\NIM(#1,#2,#3){Nucl.\ Instrum.\ Meth.\ \issue(#1,#2,#3)}
\def\PL(#1,#2,#3){Phys.\ Lett.\ \issue(#1,#2,#3)}
\def\PRD(#1,#2,#3){Phys.\ Rev.\ D \issue(#1,#2,#3)}
\def\PRL(#1,#2,#3){Phys.\ Rev.\ Lett.\ \issue(#1,#2,#3)}
\def\SJNP(#1,#2,#3){Sov.\ J. Nucl.\ Phys.\ \issue(#1,#2,#3)}
\def\ZPC(#1,#2,#3){Zeit.\ Phys.\ C \issue(#1,#2,#3)}

\begin{document} 

\rightline{CU-PHYSICS-10-2006}

\centerline{\Large\bf{Signals of Universal Extra Dimension at the ILC}}

\begin{center}
{\large{Biplob Bhattacherjee \footnote{Presented at the Linear Collider
Workshop (LCWS06), Bangalore, India, March 2006}}}\\
Department of Physics, Calcutta University, 92 A.P.C. Road,
Kolkata - 700009, India
\end{center}
\abstract{
In the minimal Universal Extra Dimension model, single production of $n=2$
gauge bosons provides a unique discriminating feature from supersymmetry.
We discuss how the proposed International Linear Collider can act as
a $n=2$ factory, much in the same vein as LEP.}


%
%
%
%


\section{Introduction}

In the simplest Universal Extra Dimension (UED) model
proposed by Appelquist, Cheng, and Dobrescu \cite{acd}, 
there is only one
extra dimension, denoted by $y$, compactified on a circle ($S_1$) of
radius $R$. All SM particles can access this dimension.
To get chiral fermions at low-energy, one must impose a further $\ztwo$
symmetry ($y\leftrightarrow -y$), so that finally we have an $S_1/\ztwo$
orbifold. 
As is well-known, a higher dimension theory is nonrenormalisable
and should be treated in the spirit of an effective theory  valid upto
a scale $\Lambda > R^{-1}$. All fields have five space-time components;
when brought down to four dimensions, for each low-mass (zero-mode) Standard
Model (SM) particle of mass $m_0$, 
we get an associated Kaluza-Klein (KK) tower, the
$n$-th level (this $n$ is the KK number of the particle)
of which has a mass given by $ m_n^2 = m_0^2 + \frac{n^2}{R^2}.$
This is a tree-level relationship and gets modified once we take into
account the radiative corrections.
The KK-number is conserved in the tree-level theory; this means that the 
lowest-mass $n=1$ particle,
which turns out to be the $n=1$ photon, is absolutely stable. Such a
lightest KK particle (LKP), just like the lightest supersymmetric particle
(LSP), is an excellent candidate for dark matter.

Radiative corrections to the masses of the KK particles have been computed
in \cite{georgi,cheng1,pk}. These papers, in particular \cite{cheng1}, show
that the almost mass-degenerate spectrum for any KK level
splits up due to such correction terms. There are two
types of correction; the first one, which results just from the compactification
of the extra dimension, is in general small (zero for fermions) and is
constant for all $n$ levels. This we will call the bulk correction. The second
one, which we will call boundary correction, is comparatively large (goes
as $\ln\Lambda^2$ and hence, in principle, can be divergent), and plays
the major role in  determining the exact spectrum and possible decay modes.
The boundary correction terms are related with the interactions present only
at the fixed points $y=0$ and $y=\pi R$. If the interaction is symmetric 
under the exchange of these fixed points (this is another $\ztwo$ symmetry,
but not the $\ztwo$ of $y\leftrightarrow -y$), the conservation of KK
number breaks down to the conservation of KK parity, defined as $(-1)^n$.
Thus, LKP is still stable, but it is possible to produce an $n=2$ state
from two $n=0$ states. This particular feature will be of central interest
to this talk.
With the proposed reach of ILC in mind, we will focus on the range 
300 GeV $<R^{-1}<$ 500 GeV. A more detailed discussion and relevant
references can be found in \cite{bk01,bk02}. 


Let us mention here that though
the main focus is on the ILC, an identical study may be performed for
CLIC.  Clearly, the reach of CLIC will be much higher.

It has been pointed out \cite{cheng2} that a `smoking gun' signal of
UED would be the production of $n=2$ states. Pair production of such states
is difficult even at the LHC energy, and is surely out of reach for ILC.
However, one can produce a single $\g_2$ or $Z_2$. 
These will be narrow peaks, closely spaced, and probably not resolvable at LHC. 
Here ILC will
perform a much better job, and if it can sit on these resonances, it may
even repeat the LEP-I story. Such precision measurements will definitely 
determine the model parameters, even if it is not the simplest UED model.  
There are a couple of points that the reader should note.
\begin{itemize}
\item
If a collider is energetic enough to pair produce $n=1$ excitations, single
production of $n=2$ states is also possible. Since it is not possible to
produce only one $n=1$ UED state, it is a none-or-both situation.  
\item 
Decay of a $n=2$ state to two $n=0$ states is allowed by KK parity conservation,
but this is suppressed by boundary-to-bulk ratio. However, there is no
phase space suppression, not even if the final state is a $t\bar{t}$ pair.
On the other hand, the coupling is large for the KK number conserving decays
($2\to 2-0,1-1$, where the numbers are for the generic KK levels), 
but there is a heavy kinematic suppression. Ultimately
it turns out that both suppressions are of equal importance \cite{cheng1} and
hence both KK conserving and KK violating decays are to be taken into
account.
\end{itemize} 

In this talk we will discuss the role that ILC may play in studying this
resonance physics. 

\section{The KK number violating interactions}

A consistent formulation of UED needs the inclusion of interaction terms that
exist only at the fixed points \cite{georgi,cheng1}. 
In the simplest UED model, these terms are taken to be universal, 
symmetric about
the fixed points, and vanishing for energy $\Lambda \gg R^{-1}$. This
introduces only two new parameters in the model, $\Lambda$ and $R^{-1}$,
and ensures the conservation of KK parity. 
(In fact, there is a third
parameter, ${\bar{m}_h}^2$, the Higgs mass term induced on the fixed
points. In the minimal UED model this is assumed to be zero, but its
precise value may be probed through a precision study \cite{bk02}.) 


The excited states of $Z$ and photon are obtained by diagonalising the 
mass matrix of $W_3$ and $B$. It has been shown in \cite{cheng1} that
for all practical purpose, the $n=2$ excitation of $Z$ is almost $W_3$
(so that it is a pure SU(2) triplet and couples only to the left-handed
fermions) while the $n=2$ excitation for photon is almost a pure $B$
(so that it couples with different strengths to left- and right-handed 
fermions). 

We will be interested in the coupling of $n=2$ gauge bosons with an
$n=0$ fermion-antifermion pair. This coupling is given by \cite{cheng1}
\be
\left(-ig\g^\mu T_a P_+\right) \frac{\stwo}{2} \left(
\frac{\bar\delta(m_{V_2}^2)}{m_2^2}-2\frac{\bar\delta(m_{f_2})}{m_2}\right),
   \label{defx}
\ee
where $g$ is the generic gauge coupling, $T_a$ is the group generator
(third component of isospin, or hypercharge), and $P_+$ is the 
$\ztwo$-even projection operator, which is $P_L=(1-\g_5)/2$ for $Z_2$,
but can be both $P_L$ or $P_R$ for $\g_2$. $V$ can be either $Z$ or $\g$.
The expressions for the
boundary corrections, $\bar\delta$, can be found in \cite{cheng1}.  

\begin{figure}
\vspace{-10pt}
\centerline{\hspace{-3.3mm}
\rotatebox{-90}{\epsfxsize=6cm\epsfbox{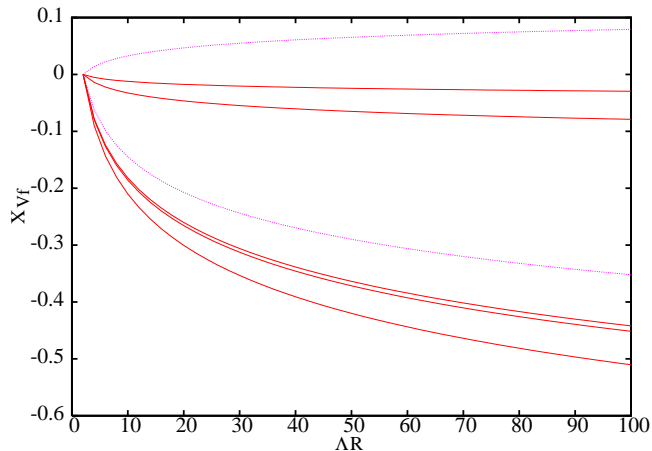}}}
\caption{$X_{Vf}$, the KK number violating couplings, as
a function of $\Lambda R$, for $R^{-1}=300$ GeV (the values are independent
of $R$). From top to bottom, the curves are for $X_{ZL}$, $X_{\g e}$,
$X_{\g L}$, $X_{ZQ}$, $X_{\g d}$, $X_{\g u}$, and $X_{\g Q}$ respectively.
For their definitions, see text.}
\end{figure}

It is easy to check that for any level, the excitation of the photon,
$\g_n$, is the lowest-lying particle. Thus, $\g_2$ cannot decay into a pair
of $n=0$ and $n=2$ fermions. In fact, the decay to an $n=1$ pair is also
kinematically forbidden, for all choices of $\Lambda$ and $R$. Thus, the only
possible way to decay is to an $n=0$ fermion-antifermion pair. Here, both
right- and left-handed pairs (of quarks and leptons, including neutrinos)
are included, albeit with different strengths, as obtained from eq.\ 
(\ref{defx}). In figure 1, we show how the function $X_{Vf}$, defined as
\be
X_{Vf}= \frac{\stwo}{2} \left(
\frac{\bar\delta(m_{V_2}^2)}{m_2^2}-2\frac{\bar\delta(m_{f_2})}{m_2}\right),
   \label{xvfdef}
\ee
varies for $V=\g,Z$ and $f=u_i,d_i,e_i$ (SU(2) singlet states) and $L_i,
Q_i$ (SU(2) doublet states), where $i$ is the generation index. It is
obvious that $\g_2$ should decay almost entirely to a $q\bar{q}$ pair, 
because of the larger splitting between $\g_2$ and $n=2$ quarks. Altogether,
there are 45 channels, including the colour degrees of freedom. 


The decay pattern of $Z_2$ is more complicated. It is an almost pure
$(W_3)_2$, so it couples only to left-handed doublet fermions. Kinematically,
decay to an $n=1$ pair of lepton doublet ($\ztwo$-even)
is allowed, except for very low
values of $\Lambda$ ($\Lambda R < 3$). There are 6 such channels, including
neutrinos. These states will ultimately decay
to the corresponding $n=0$ leptons, plus $\g_1$, the LKP, (even the $n=1$ 
neutrino can decay in this channel), so that the signature
will be a pair of soft leptons (for charged lepton channels) 
plus a huge missing energy (excited neutrinos, of course, will go undetected).
Fortunately, these final soft leptons should be detectable \cite{bdkr,asesh}. 
Similarly, $Z_2$ can decay to a pair of $n=2$ and $n=0$ doublet leptons.
Again, there are 6 channels, plus 6 CP-conjugate ones.
Both these modes are KK-number conserving, but there is an important
difference: while the coupling is the usual $g$ for the latter channels,
it is $g/\stwo$ for the former ones. This can be checked by integrating
the trigonometric terms dependent on the fifth coordinate $y$. 

Just like $\g_2$, $Z_2$ has its own share of KK-number violating modes,
but it can only decay to a left-handed pair. Since the lower limit on $R^{-1}$
is about 300 GeV, both these gauge bosons can decay even to the $n=0$
$t\bar{t}$ pair. However, KK-number conserving $Z_2$ decays to electroweak
bosons are forbidden from kinematic considerations.


In the minimal UED model, ${\bar{m}_h}^2=0$, $Z_2$ cannot decay through
the Bjorken channel to $Z_1 h_1$, purely from kinematic considerations.
(The three-body channels, with a virtual $Z_1$ or $h_1$, will be even
more suppressed.)
However, if ${\bar{m}_h}^2 < 0$, all the Higgs masses will be lowered,
and one can just be able to produce a neutral CP-even Higgs excitation
through this channel. The decay channel of $h_1$ is dominantly a right-handed
$\tau$ pair (assuming the mixing in the $n=1$ level to be small) plus
LKP, and if the $\tau$s are soft enough, they may escape detection,
leading to an invisible decay mode of $h_1$. Of course, the vertex
$Z_2 W_1^\pm h_1^\mp$ does not exist.

\section{Production and decay of n = 2 neutral gauge bosons}

The gauge bosons are produced as $s$-channel resonances in $e^+e^-$
collision through KK-number violating couplings. This suppression brings
down the peak cross-section to an otherwise expected nanobarn level to about
35-45 pb for $Z_2$ and about 63 pb for $\g_2$ (for $R^{-1}=300$ GeV,
and the variation is due to that of $\Lambda$). For $R^{-1}=450$ GeV,
these numbers drop to 16-21 pb and 28 pb, respectively. The reason for
a higher production cross-section for $\g_2$ is its narrower width compared
to $Z_2$. However, it will be almost impossible to detect $\g_2$ at LHC 
since it decays almost entirely to two jets which will be swamped by the
QCD background, and moreover the resonance is quite narrow. $Z_2$ has a
better chance, since there are a number of hadronically quiet channels, and
soft leptons with energy greater than 2 GeV should be detectable. But for
a precision study of these resonances we must turn to ILC (or CLIC). These
machines should be able to measure precisely the positions and the widths
of these two peaks, and hence entirely determine the spectrum, since there
are only two unknown parameters (hopefully the Higgs mass will already
be measured by LHC). These measurements, in conjunction with the precise
determination of $n=1$ levels, should be able to discriminate, not only
between UED and supersymmetry, but even the minimal version of UED from its
variants.

\begin{figure}
\vspace{-10pt}
\centerline{\hspace{-3.3mm}
\rotatebox{-90}{\epsfxsize=6cm\epsfbox{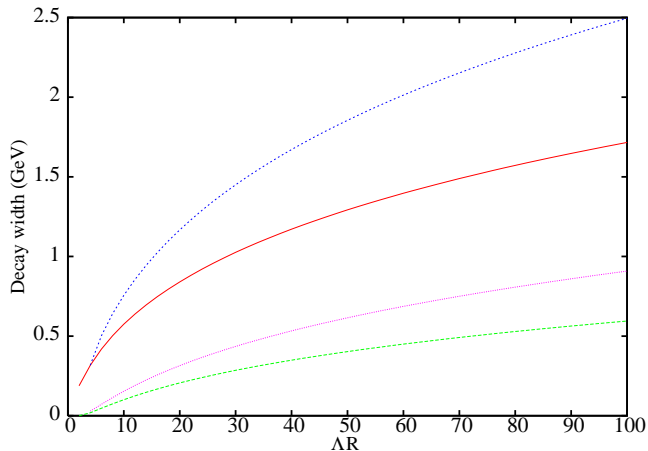}}}
\caption{Decay widths of $Z_2$ (upper pair) and $\g_2$ (lower pair)
as a function of $\Lambda R$, for $R^{-1}=450$ GeV and 300 GeV (upper and 
lower curves in a pair).}
\end{figure}

In figure 2 we show the decay widths of $Z_2$ and $\g_2$, plotted for
two different values of $R^{-1}$ and as a function of $\Lambda R$. They
increase logarithmically, because of the $\log\Lambda^2$ dependence of
the couplings, but no new channel opens up. For small values of $\Lambda R$
(2-3), the KK-number conserving channels for $Z_2$ are still closed, and 
$Z_2$ can be very long-lived, even to leave a displaced vertex. (As discussed
earlier, for $\Lambda R=2$, a somewhat fine-tuned value, $Z_2$ is almost
stable, and the peak is correspondingly narrow and hence difficult to
detect.)

We emphasize that this study will be meaningful only if LHC finds some
signal of new physics, which may look like UED, and for which the pair
production of $n=1$ states is not beyond the reach of ILC. In that case
a careful scan about $\sqrt{s}=2/R$ should reveal these two peaks. The points
that one would like to verify are:\\
(i) On the $Z_2$ peak, ${\cal R}$, the ratio of $e^+e^-$ to two jets to
$e^+e^-\to\mu^+\mu^-$ would show a sharp dip, in particular if we include
the missing energy events. The reason is that the $Z_2$-width is dominated by
the channel to a pair of $n=1$ leptons, and quarks can appear only from 
KK-number violating interactions. On the other hand, ${\cal R}$ should show
a sharp peak on the $\g_2$ resonance.\\
(ii) The cross-section would show a kink between the two peaks; this is
the position where the KK-number conserving channels open up.\\
(iii) With the polarised beam option, the behaviour of the two peaks will
be quite different. Since $Z_2$ couples only to the left-handed fermions,
with suitable polarisation the peak may vanish altogether, or may get
enhanced by a factor of 3 (assuming 80\% $e^-$ polarisation and 60-70\%
$e^+$ polarisation). The $\g_2$ peak will get enhanced by about a factor
of 2 with left-polarised $e^-$ beam, but will never vanish altogether.

Let us also note that the SM background, coming from the continuum, is
less than 10 pb for $\sqrt{s}=600$-900 GeV \cite{jlc}, and may be
further reduced by suitable cuts.

%

\end{document}